\documentclass[aps,prl,twocolumn,superscriptaddress,longbibliography]{revtex4-2}

\usepackage{gensymb}
\usepackage{amssymb}

\usepackage{color}
\usepackage[normalem]{ulem}
\usepackage{graphicx}
\usepackage{dcolumn}
\usepackage{float}
\usepackage{bm}
\usepackage{amsmath}
\usepackage{comment}

\newcommand{\be}{\begin{equation}}
\newcommand{\ee}{\end{equation}}
\newcommand{\bfig}{\begin{figure}}
\newcommand{\efig}{\end{figure}}

\newcommand{\HT}{HoTe$_3$}
\newcommand{\RT}{$R$Te$_3$}
\newcommand{\TT}{TbTe$_3$}
\newcommand{\GT}{GdTe$_3$}
\newcommand{\DT}{DyTe$_3$}

\usepackage{multirow}
\usepackage{makecell}
\makegapedcells

\begin{document}

\title{Stripe antiferromagnetism in van der Waals metal \HT{} \\decoupled from charge density wave order}

\author{Weiyi Yun}
\affiliation{Department of Applied Physics and Quantum-Phase Electronics Center (QPEC), The University of Tokyo, Bunkyo, Tokyo 113-8656, Japan}

\author{Ryota Nakano}
\affiliation{Department of Applied Physics and Quantum-Phase Electronics Center (QPEC), The University of Tokyo, Bunkyo, Tokyo 113-8656, Japan}

\author{Ryo Misawa}
\affiliation{Department of Applied Physics and Quantum-Phase Electronics Center (QPEC), The University of Tokyo, Bunkyo, Tokyo 113-8656, Japan}

\author{Rinsuke Yamada}
\affiliation{Department of Applied Physics and Quantum-Phase Electronics Center (QPEC), The University of Tokyo, Bunkyo, Tokyo 113-8656, Japan}

\author{Shun Akatsuka}
\affiliation{Department of Applied Physics and Quantum-Phase Electronics Center (QPEC), The University of Tokyo, Bunkyo, Tokyo 113-8656, Japan}

\author{Yoshichika \={O}nuki}
\affiliation{RIKEN Center for Emergent Matter Science (CEMS), Wako, Saitama 351-0198, Japan}

\author{Priya Ranjan Baral}
\affiliation{Department of Applied Physics and Quantum-Phase Electronics Center (QPEC), The University of Tokyo, Bunkyo, Tokyo 113-8656, Japan}

\author{Sebastian Esser}
\affiliation{Department of Applied Physics and Quantum-Phase Electronics Center (QPEC), The University of Tokyo, Bunkyo, Tokyo 113-8656, Japan}

\author{Hiraku Saitoh}
\affiliation{The Institute for Solid State Physics, The University of Tokyo, Kashiwa 277-8581, Japan}

\author{Ryoji Kiyanagi}
\affiliation{J-PARC Center, Japan Atomic Energy Agency, Tokai 319-1106, Japan}

\author{Takashi Ohhara}
\affiliation{J-PARC Center, Japan Atomic Energy Agency, Tokai 319-1106, Japan}

\author{Taro Nakajima}
\affiliation{The Institute for Solid State Physics, The University of Tokyo, Kashiwa 277-8581, Japan}

\author{Taka-hisa Arima}
\affiliation{RIKEN Center for Emergent Matter Science (CEMS), Wako, Saitama 351-0198, Japan}
\affiliation{Department of Advanced Materials Science, The University of Tokyo, Kashiwa, Chiba 277-8561, Japan}

\author{Max Hirschberger}
\email{hirschberger@ap.t.u-tokyo.ac.jp}
\affiliation{Department of Applied Physics and Quantum-Phase Electronics Center (QPEC), The University of Tokyo, Bunkyo, Tokyo 113-8656, Japan}
\affiliation{RIKEN Center for Emergent Matter Science (CEMS), Wako, Saitama 351-0198, Japan}

\date{\today}

\begin{abstract}
The $R\mathrm{Te}_3$ ($R = \text{rare earth}$) family of layered van der Waals (vdW) compounds hosts coexisting magnetic and charge density wave (CDW) orders, yet the interplay between these degrees of freedom remains largely unexplored. Combining polarized and unpolarized neutron diffraction on single crystals of $\mathrm{HoTe}_3$, we identify two distinct antiferromagnetic (AFM) phases, both exhibiting a collinear $\uparrow\uparrow\downarrow\downarrow$ motif within individual vdW layers.
The two phases are distinguished by the vdW stacking of magnetic layers: ferromagnetic (FM) stacking in the higher-temperature AFM-II phase, here termed ``vertical-stripe'', and AFM stacking in the AFM-I ground state, here termed ``tilted-stripe''; the two phases have propagation vectors $\bm{q}_{\mathrm{m2}} = (0.48, 0, 0)$ and $\bm{q}_{\mathrm{m1}} = (0.5, 0.5, 0)$, respectively.
In contrast to the CDW-driven exotic magnetism in $\mathrm{DyTe}_3$, $\mathrm{TbTe}_3$, and $\mathrm{GdTe}_3$, we find no evidence for coupling between magnetism and CDW in $\mathrm{HoTe}_3$. The presence of a checkerboard-type charge order in \HT{} may be detrimental for realizing coupled spin/charge order parameters in layered vdW systems. 
\end{abstract}

\maketitle
\begin{figure}[htb]
 \centering
  \includegraphics[width=1\linewidth]{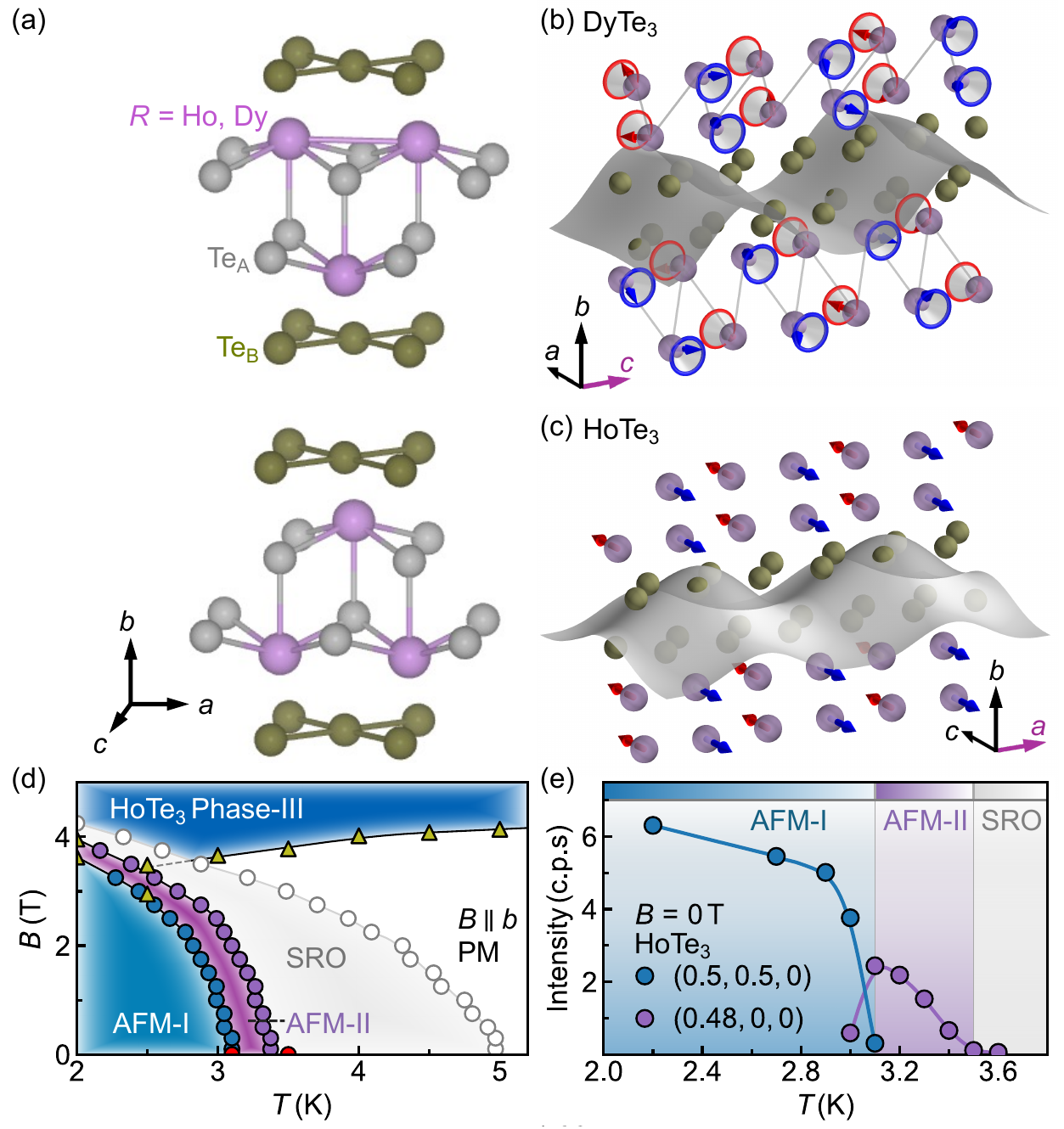}
\caption{
(color online). Magnetism in \RT{} ($R=$ Ho, Dy) with charge density wave (CDW) order. 
(a) Orthorhombic crystal structure (space group \textit{Cmcm}, No.~63) in layered \RT{}, with covalently bonded $R$--Te$_\mathrm{A}$ bilayers and conducting Te$_\mathrm{B}$ square nets~\cite{footnote1}. The $b$-axis is perpendicular to the vdW layers.
(b) \DT{}'s CDW is predominantly unidirectional along $c$.
Its conical magnetic texture is modulated along the $c$-axis through coupling to the CDW. (c) \HT{}'s CDW has stronger checkerboard character, with charge modulations along two orthogonal directions.
Its AFM-I ground state has moments parallel to the $c$-axis and is independent of the checkerboard CDW. 
Te$_\mathrm{A}$ atoms and CDW-induced lattice distortions are not depicted. 
(d) Magnetic phase diagram of \HT{} with $B$ applied along the $b$-axis. Phase boundaries are from magnetization measurements, but the red circles ($B = 0$) correspond to the onset of magnetic Bragg intensities in panel (e). 
The light gray region between the PM and AFM-II phases is thought to host short-range magnetic order (SRO). (e) Temperature dependence of magnetic reflection intensities in \HT{} corresponding to the AFM-I and AFM-II propagation vectors.
}
    \label{Fig1}
\end{figure}

\textit{Introduction}~--~Rare-earth tritellurides (\RT{}) are quasi-two-dimensional, layered van der Waals (vdW) compounds that exhibit rich electronic and magnetic phenomena. The \RT{} are known to host robust charge density waves (CDWs)~\cite{brouet2004fermi,walmsley2020magnetic,kogar2020light,rettig2016persistent,schmitt2008transient,chikina2023charge,yumigeta2021advances,PhysRevB.90.085105,PhysRevB.77.235104,ru2008magnetic,PhysRevB.77.035114,seong2021angle,malliakas2005square} related to Fermi surface nesting between nearly parallel sheets derived from Te-$5p$ orbitals~\cite{PhysRevB.81.073102,PhysRevB.77.035114,PhysRevB.77.235104,kang2024distinct,Lee2016PrTe3}. In \RT{} with heavy rare earths, the CDW order is a checkerboard pattern or superposition of two charge modulations with orthogonal propagation directions along the crystallographic $c$- and $a$-axes, respectively. Figure~\ref{Fig1}(c) depicts this complex checkerboard CDW as a grey shaded surface. The checkerboard CDW is highly sensitive to the lattice volume as controlled by the rare earth $R$ and yields to unidirectional CDW at high temperature or smaller $R$~\cite{Ru2008cdw,Pfuner2010}.
Antiferromagnetic order~\cite{liu2020electronic,lei2020high,raghavan2024atomic,deguchi2009magnetic} is present in the \RT{} at low temperature, and superconductivity can be induced via external pressure~\cite{hamlin2009pressure,PhysRevB.91.205114}. The intricate coupling among charge, orbital, lattice and spin degrees of freedom makes \RT{} an ideal platform for exploring correlation phenomena in low-dimensional systems. It is also notable that atomically flat surfaces of \RT{} can be prepared by mechanical exfoliation, for advanced experimental techniques. There is thus a prospect for vdW heterostructure engineering using this class of materials~\cite{lei2020high}.

$R\mathrm{Te}_3$ compounds crystallize in an orthorhombic structure (space group $Cmcm$, No.~63), as shown in Fig.~\ref{Fig1}(a). The crystal lattice consists of bilayers of highly metallic Te square nets separated by covalently bonded $R$--Te slabs, with the stacking direction corresponding to the orthorhombic $b$-axis. Here, we focus on two representative members of the family: Fig.~\ref{Fig1}(b) shows the coupled spin and charge order of \DT{}, with a predominantly unidirectional CDW~\cite{footnote2}. A collinear $\uparrow\uparrow\downarrow\downarrow$ magnetic structure couples to the CDW, stabilizing an exotic helimagnetic cone phase~\cite{akatsuka2024non}. Similar coupling between CDW and AFM order has been reported in \GT{}~\cite{lei2020high} and \TT{}~\cite{shamova2025comparative,chillal2020strongly}, but the detailed spin patterns of \TT{} and \GT{} remain to be solved.


In this Letter, we combine neutron diffraction and magnetization measurements to determine the magnetic phases and spin textures of \HT{} in presence of a CDW with strong checkerboard character. We identify collinear $\uparrow\uparrow\downarrow\downarrow$ type AFM order within individual vdW layers, as depicted in Fig.~\ref{Fig1}(c). 
Our magnetic structure analysis is based on symmetry considerations, supported by polarized neutron diffraction, as well as full refinement of the unpolarized magnetic neutron scattering intensities. Since no evidence is found for cross-talk between CDW and AFM order in \HT{}, we suggest that the checkerboard CDW is detrimental for coupling charge and spin order in this class of materials, contrary to the (predominantly) unidirectional CDW. 
\bigskip

\begin{figure}[htb]
  \centering
  \includegraphics[width=1\linewidth]{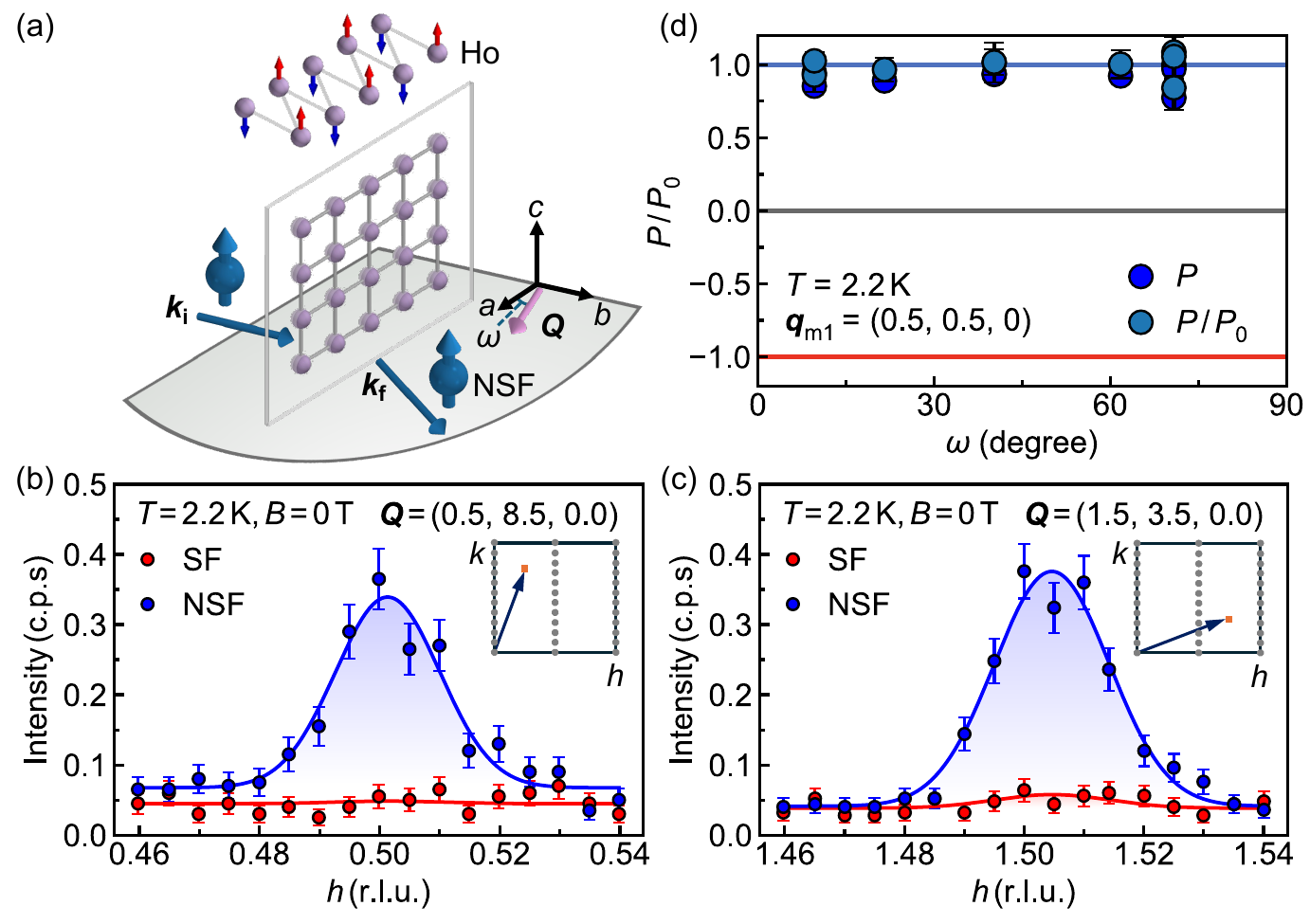}
\caption{
(color online). 
Polarized neutron scattering in phase AFM-I of \HT{}. 
(a) The incoming and outgoing wavevectors of the neutron beam, $\bm{k}_\mathrm{i}$ and $\bm{k}_\mathrm{f}$, span the $hk0$ scattering plane (gray). The total momentum transfer is $\bm{Q}$. In the Ho sublattice, moments are collinear along the $c$-axis (perpendicular to the scattering plane), yielding finite non-spin flip (NSF) and zero spin-flip (SF) scattering intensity. Spin directions of polarized neutrons are shown as blue balls with arrows. (b,c) $h$-scans at $\bm{Q}=(0.5,8.5,0)$ and $(1.5,3.5,0)$. The NSF and SF intensities are fitted with Gaussian functions. The small SF signal arises from imperfect spin-flipper efficiency; within experimental uncertainty, the scattering is entirely NSF.
Insets: schematic of $\bm Q$ in the scattering plane.
(d) Flipping ratio $P = (I_\mathrm{NSF}-I_\mathrm{SF})/(I_\mathrm{NSF}+I_\mathrm{SF})$ at $T = 2.2$ K for magnetic reflections lying in the scattering plane $hk0$, measured as a function of $\omega$, the angle between each reflection's momentum transfer $\bm{Q}$ and the $a$-axis ($h00$). $P$ is normalized to $P_0$ to account for imperfect beam polarization (see text). When all points are on the red (blue) line, the moments are fully parallel (fully perpendicular) to the scattering plane.
}
  \label{Fig2}
\end{figure}

\begin{figure}[htb]
 \centering
  \includegraphics[width=1\linewidth]{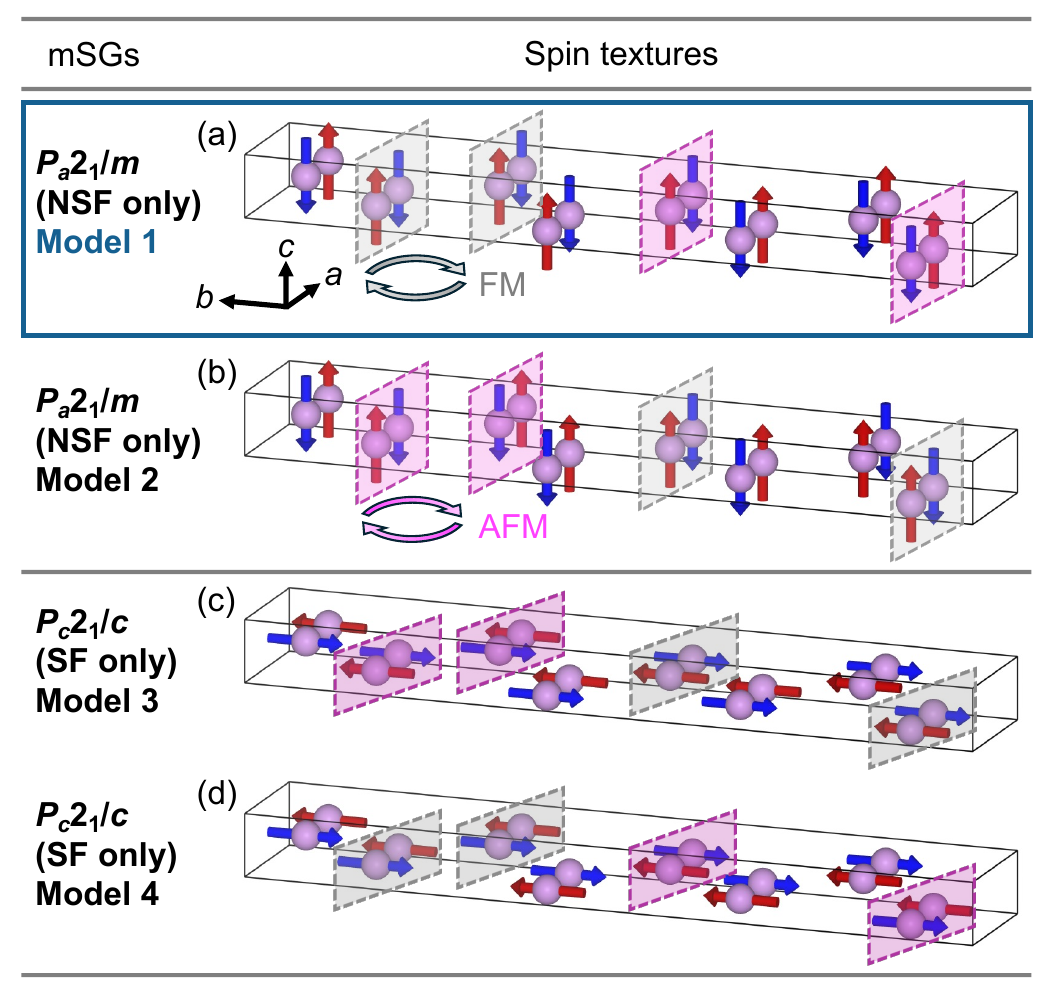} 
  \caption{
(color online). Symmetry analysis for phase AFM-I. We show the highest-symmetry magnetic subgroups (mSGs) of the paramagnetic space group $Cmcm1'$ and the corresponding candidate spin textures. (a,b) Magnetic structures in the $P_a2_1/m$ mSG exhibit Ising-like collinear AFM order along the $c$-axis, consistent with polarized neutron scattering in Fig.~\ref{Fig2}. (c,d) Magnetic structures in the $P_c2_1/c$ mSG have their moments perpendicular to $c$, inconsistent with polarized neutron scattering. In each mSG, the Ho atoms at the Wyckoff $4c$ positions are divided into two sets, termed inner and outer set, which are separated by interlayer spacings of approximately $b/3$ and $2b/3$, respectively. In Model~1 (Model~2), the outer (magenta) and inner (gray) Ho sets are AFM  (FM) and FM (AFM) coupled, respectively. The structures (c) and (d) can be discussed on the same grounds. Only Model~1 [blue box in (a)] is consistent with the analysis in Fig.~\ref{Fig4}.
    }
    \label{Fig3}
\end{figure}

\begin{figure*}[htb]
  \centering
  \includegraphics[width=1\linewidth]{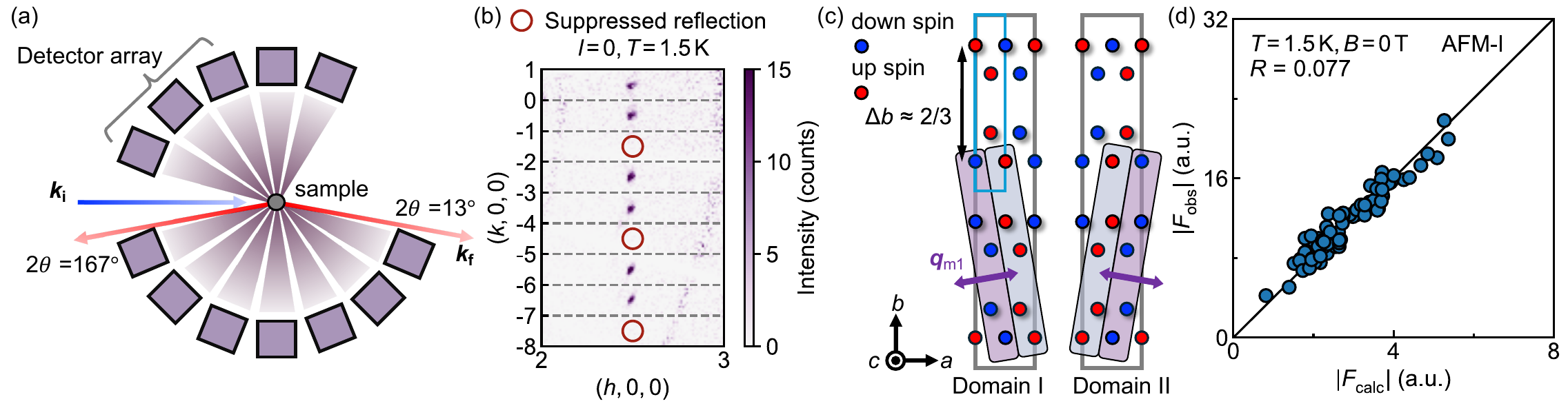} 
\caption{%
(color online) 
Refinement of the magnetic structure in phase AFM-I of \HT{}. 
(a) Geometry of SENJU, a time-of-flight (TOF) Laue diffractometer at J-PARC. Red and blue arrows indicate the incoming and outgoing neutron wavevectors, $\bm k_\mathrm{i}$ and $\bm k_\mathrm{f}$. 
(b) Reconstructed reciprocal space map in the $hk0$ plane at $T = 1.5\,$K. In AFM-I, there is a pronounced suppression of intensity at $k = -3/2, -9/2, -15/2$ when $l=0$, consistent with Eq.~\eqref{extinction}. 
(c) Two domains of Model~1 for AFM-I, showing 
alternating tilted-stripe spin textures. Only Ho sites are shown. The gray box indicates the conventional magnetic unit cell, four times larger than the conventional crystallographic unit cell (blue box). Moments align along the $c$-axis with propagation vector $\bm{q}_{\mathrm{m1}}$ (purple arrow). The tilted grey and violet boxes with round edges are guides to the eye, indicating planes of parallel spins -- i.e., spin stripes. 
(d) Magnetic refinement of unpolarized neutron intensity using Model~1. $F_\mathrm{obs}$, $F_\mathrm{calc}$ and $R$ denote observed and calculated magnetic structure factors, as well as the agreement factor of the refinement.
}
  \label{Fig4}
\end{figure*}
\textit{Results}~--~The magnetic phase diagram of \HT{} in the $B$--$T$ plane [Fig.~\ref{Fig1}(d)] reveals two successive antiferromagnetic phases, AFM-I and AFM-II. The zero-field phase transition between these phases is clearly visible in the temperature dependence of the magnetic scattering intensity [Fig.~\ref{Fig1}(e)], where two magnetic propagation vectors $\bm q$ are identified. 
We focus on the ground state, AFM-I, in the following, and study its magnetic structure first by polarized neutron scattering.
The experimental geometry is shown in Fig.~\ref{Fig2}(a)~\cite{Nakajima2024PONTA}. We align the neutron spin quantization direction along the crystallographic $c$-axis, with the scattering plane defined as $hk0$.  
Neutron scattering is sensitive only to $\bm{M}^{\perp}(\bm{Q})$, the magnetization component perpendicular to the total momentum transfer $\bm{Q}$. 
In our geometry, the non-spin-flip (NSF) and spin-flip (SF) channels selectively probe components of $\bm{M}^{\perp}(\bm{Q})$ parallel and perpendicular to the $c$-axis.
Representative $h$-scans across magnetic Bragg peaks [Fig.~\ref{Fig2}(b,c)] show intensity exclusively in the NSF channel. The absence of SF scattering at all observed reflections is quantified by the flipping ratio $P$, shown and defined in Fig.~\ref{Fig2}(d). 
The normalized flipping ratio $P/P_0$ for all observed reflections in the $hk0$ scattering plane is close to unity, 
indicating that the Ho$^{3+}$ moments are fully aligned along the $c$-axis. Here, $P_0$ is the reference flipping ratio measured on a strong nuclear (non-magnetic) reflection to account for the imperfect polarization of the incident neutron beam and the analyzer efficiency. We therefore identify the AFM-I phase in HoTe$_3$ as an Ising-like collinear antiferromagnet with moments aligned along the $c$-axis.

We use symmetry analysis to identify the most likely magnetic structure models for the AFM-I phase of \HT{}, before comparing these models to unpolarized neutron diffraction data. 
We determine the highest symmetry subgroup of the paramagnetic space group $Cmcm1'$, i.e., the maximal magnetic subgroup. The analysis is constrained by (i) the $\mathrm{Ho}^{3+}$ ions occupying the $4c$ Wyckoff positions and (ii) the magnetic propagation vector $\bm{q}_{\mathrm{m1}}=(0.5,0.5,0)$, which results in two candidate mSGs of high symmetry, where each mSG provides two non-identical magnetic structure models (Fig.~\ref{Fig3}). The candidate mSGs differ in their magnetic anisotropy: $P_a2_1/m$ enforces an Ising-like magnetic anisotropy with moments aligned along the $c$-axis, whereas $P_c2_1/c$ requires an XY-like magnetic anisotropy with moments perpendicular to the $c$-axis. Polarized neutron scattering uniquely selects $P_a2_1/m$ as the correct mSG, since magnetic reflections with $\bm{Q}$ in the $hk0$ plane appear exclusively in the NSF channel. Consequently, two symmetry-allowed structure models within $P_a2_1/m$ remain, as illustrated in Fig.~\ref{Fig3}(a,b).

To distinguish between Model~1 and Model~2 in Fig.~\ref{Fig3}, and also to determine the magnetic domain populations, we performed time-of-flight Laue neutron diffraction using the SENJU diffractometer at J-PARC [Fig.~\ref{Fig4}(a)]. The detector pattern in phase AFM-I exhibits magnetic superlattice reflections [Fig.~\ref{Fig4}(b)] with a pronounced suppression of intensity along the $(0k0)$ direction. This quasi-extinction behavior places a stringent constraint on the magnetic structure factor. Taking into account scattering vectors of the form
$\bm{Q} = (h',k',l) = \bm{G} + \bm{q}_{\mathrm{m1}}$, where $\bm{G}=(h,k,l)$ is a reciprocal lattice vector, 
the structure factor for Model~1 shown in Fig.~\ref{Fig3}(a) 
is evaluated following Supplementary Note~6~\cite{SI}. 
We find that the suppressed reflections satisfy
\begin{equation}
k' = k + \frac{1}{2} = \frac{3n}{2}, 
\quad n \in \mathbb{Z},\; n~\text{odd},
\label{extinction}
\end{equation}
when $l = 0$. For example, the intensity is predicted to be suppressed at $k = \pm 3/2, \pm 9/2,...$ as observed experimentally in Fig.~\ref{Fig4}(b). 
Model~2 cannot explain these suppressed intensity patterns~\cite{SI}, and we thus conclude that the AFM-I phase in \HT{} 
exhibits a tilted-stripe antiferromagnetic structure corresponding to Model~1, illustrated 
in Fig.~\ref{Fig4}(c). The construction of AFM-I's spin pattern based on symmetry 
is detailed in Supplementary Note~9~\cite{SI}.

To quantitatively assess the consistency between the proposed model and the experimental data, we performed magnetic structure refinement,
treating the relative population of domains depicted in Fig.~\ref{Fig4}(c) as the sole adjustable parameter~\cite{SI}. Figure~\ref{Fig4}(d) shows the optimal refinement, which produces a population ratio of $48.1:51.9\,\%$ between two magnetic domains related by a mirror plane perpendicular to the $a$-axis.
The refined Ho$^{3+}$ moment, $7.89~\mu_\mathrm{B}$ per holmium, is reduced from the free-ion value of $10~\mu_\mathrm{B}/\mathrm{Ho}^{3+}$. We note that, at the experimental $T = 1.5~\mathrm{K}$ ($T/T_{N1} \approx 0.48$), the system remains relatively close to the critical temperature. We expect the ordered moment to approach the free-ion value at zero temperature.

For phase AFM-II, polarized neutron scattering experiments, detailed in the Supplemental Material, again show a flipping ratio $P/P_0=1$ for reflections in the $hk0$ plane, indicating that the Ho moments remain collinear along the crystallographic $c$-axis even at elevated $T$. The magnetic propagation vector is $\bm{q}_{\mathrm{m2}} = (0.48, 0, 0)$, with a slight deviation from $(1/2, 0, 0)$, which suggests that a magnetic discommensuration occurs roughly every $50$ unit cells along the $a$-axis~\cite{SI}. Following a similar symmetry analysis as for AFM-I, we identify two candidate mSGs for AFM-II, $P_c nma$ and $P_a bcm$, both of which preserve the $c$-mirror symmetry. Our magnetic structure refinement identifies $P_c nma$ as the mSG of the AFM-II phase, while the alternative $P_a bcm$ fails to reproduce the observed intensities \cite{SI}. 
Under the symmetry constraints imposed by the $P_c nma$ magnetic space group, the AFM-II spin texture adopts an Ising-like vertical-stripe pattern, with $\uparrow\uparrow\downarrow\downarrow$ patterns in single layers and ferroic stacking along the $b$-axis~\cite{SI}. 

\textit{Conclusions}~--~Polarized neutron scattering reveals the magnetic symmetry of the commensurate AFM-I phase in \HT{} to be the magnetic space group $P_a2_1/m$. By 
magnetic structure refinements, we establish that both AFM-I and AFM-II have $\uparrow\uparrow\downarrow\downarrow$ structures in individual layers, but they differ by their stacking along the $b$-axis.

In both phases, the closest neighbors across the metallic tellurium layer (distance about $1/3$ of the lattice constant $b$) are coupled ferromagnetically; c.f.~Fig.~\ref{Fig4}(c). In contrast, the Ho$^{3+}$ ions stacked at a distance $2b/3$ are coupled antiferromagnetically in AFM-I but ferromagnetically in AFM-II, resulting in a tilted-stripe pattern in the former and a vertical-stripe pattern in the latter case. 

In contrast to the XY-like magnetic anisotropy in $R\mathrm{Te}_3$ with light rare earths~\cite{iyeiri2003magnetic,okuma2020fermionic}, \HT{} exhibits robust Ising-like magnetic anisotropy in both AFM phases. In \DT{}, the coupled spin-charge order is manifested by the presence of coupled reflections $\bm{q}_\mathrm{CDW}\pm\bm{q}_\mathrm{AF}$ in the magnetic diffraction pattern, where $\bm{q}_\mathrm{CDW}=(0, 0, 0.3)$ and $\bm{q}_\mathrm{AF} = (0, 0, 0.5)$ are the primary propagation vectors of the charge order and the antiferromagnetic texture, respectively. 
The absence of such coupling in \HT{} is distinct from the behavior in \DT{}, \GT{} and \TT{}, pointing towards a possible role of checkerboard charge order in suppressing complex magnetic states in the \RT{}. The shift of $\bm q_{\mathrm{AF}}$ from the $c$-axis in \DT{} and to the $a$-axis in \HT{} may also contribute to decoupling the spin and charge orders in \HT{}.
Our work suggests a systematic evolution of coupled spin and charge orders across the $R\mathrm{Te}_3$ series, stemming from lanthanide-controlled volume changes to the charge density wave patterns. 
\vskip\baselineskip
\textbf{Acknowledgements}\\
P.R.~Baral was supported by the Swiss National Science Foundation (SNSF) as a Postdoc.Mobility fellow (No.~P500PT\_217697). M.H. is supported by the Deutsche Forschungsgemeinschaft (DFG, German Research Foundation) via Transregio
TRR 360 - 492547816. In this research, we used the ARIM-mdx data system~\cite{hanai2024arim}. This work was supported by JSPS KAKENHI Grant Nos. JP22F22742, JP24H01607, JP25K17336, and JP24H01604. This work was partially supported by the Japan Science and Technology Agency (JST) via CREST No.~JPMJCR20T1 (Japan), FOREST No.~JPMJFR2238, and PRESTO No.~JPMJPR259A. It was also supported by JST as part of Adopting Sustainable Partnerships for Innovative Research Ecosystem (ASPIRE), Grant No.~JPMJAP2426. This work is based on experiments performed at the Japan Research Reactor 3 and at the Materials and Life Science Experimental Facility in the Japan Proton Accelerator Research Complex (proposal Nos. 22516 and 2023B0219).
\\
\bibliography{HoTe3}

@article{Lee2016PrTe3,
  author  = {Lee, Eunsook and Kim, D. H. and Kim, Hyun Woo and Denlinger, J. D. and Kim, Heejung and Kim, Junwon and Kim, Kyoo and Min, B. I. and Min, B. H. and Kwon, Y. S. and Kang, J.-S.},
  title   = {{The $7 \times 1$ Fermi Surface Reconstruction in a Two-dimensional $f$-electron Charge Density Wave System: PrTe$_3$}},
  journal = {Sci. Rep.},
  volume  = {6},
  pages   = {30318},
  year    = {2016},
  publisher={Nature Publishing Group UK London}
}

@article{PhysRevB.81.073102,
  title = {{Fermi surface evolution across multiple charge density wave transitions in ${\text{ErTe}}_{3}$}},
  author = {Moore, R. G. and Brouet, V. and He, R. and Lu, D. H. and Ru, N. and Chu, J.-H. and Fisher, I. R. and Shen, Z.-X.},
  journal = {Phys. Rev. B},
  volume = {81},
  issue = {7},
  pages = {073102},
  numpages = {4},
  year = {2010},
  month = {Feb},
  publisher = {American Physical Society},
  doi = {10.1103/PhysRevB.81.073102},
  url = {https://link.aps.org/doi/10.1103/PhysRevB.81.073102}
}

@article{hamlin2009pressure,
  title={{Pressure-Induced Superconducting Phase in the Charge-Density-Wave Compound Terbium Tritelluride}},
  author={Hamlin, JJ and Zocco, DA and Sayles, TA and Maple, MB and Chu, J-H and Fisher, IR},
  journal={Phys. Rev. Lett.},
  volume={102},
  number={17},
  pages={177002},
  year={2009},
  publisher={APS}
}

@article{liu2020electronic,
 title={{Electronic structure of the high-mobility two-dimensional antiferromagnetic metal \GT{}}},
 author = {Liu, J. S. and Huan, S. C. and Liu, Z. H. and Liu, W. L. and Liu, Z. T. and Lu, X. L. and Huang, Z. and Jiang, Z. C. and Wang, X. and Yu, N. and Zou, Z. Q. and Guo, Y. F. and Shen, D. W.},
  journal={Phys. Rev. Mater.},
  volume={4},
  number={11},
  pages={114005},
  year={2020},
  publisher={APS}
}

@article{okuma2020fermionic, 
  title={{Fermionic order by disorder in a van der Waals antiferromagnet}}, 
  author = {Okuma, R. and Ueta, D. and Kuniyoshi, S. and Fujisawa, Y. and Smith, B. and Hsu, C. H. and Inagaki, Y. and Si, W. and Kawae, T. and Lin, H. and Chuang, F. C. and Masuda, T. and Kobayashi, R. and Okada, Y.},
  journal={Sci. Rep.}, 
  volume={10}, 
  number={1}, 
  pages={15311}, 
  year={2020}, 
  publisher={Nature Publishing Group UK London} 
}

@article{malliakas2005square,
  title={{Square Nets of Tellurium: Rare-Earth Dependent Variation in the Charge-Density Wave of RETe$_3$ (RE= Rare-Earth Element)}},
  author={Malliakas, Christos and Billinge, Simon JL and Kim, Hyun Jeong and Kanatzidis, Mercouri G},
  journal={J. Am. Chem. Soc.},
  volume={127},
  number={18},
  pages={6510--6511},
  year={2005},
  publisher={ACS Publications}
}

@misc{footnote1,
  title = {{In fact, Te$_\mathrm{B}$ includes two distinct crystallographic Wyckoff sites with same site symmetry and multiplicity.}},
}

@misc{footnote2,
  title = {{More specifically, \DT{} has a second transition around $T_\mathrm{CDW,b}=50\,$K towards a bidirectional CDW, but the checkerboard character of this order parameter is expected to be weak as compared to the predominant, unidirectional CDW.}},
}

@article{brouet2004fermi,
  title={{Fermi Surface Reconstruction in the CDW State of CeTe$_3$ Observed by Photoemission}},
  author={Brouet, V and Yang, WL and Zhou, XJ and Hussain, Z and Ru, N and Shin, KY and Fisher, IR and Shen, ZX},
  journal={Phys. Rev. Lett.},
  volume={93},
  number={12},
  pages={126405},
  year={2004},
  publisher={APS}
}

@article{kogar2020light,
  title={{Light-induced charge density wave in LaTe$_3$}},
  author = {Kogar, Anshul and Zong, Alfred and Dolgirev, Pavel E. and Shen, Xiaozhe and Straquadine, Joshua and Bie, Ya-Qing and Wang, Xirui and Rohwer, Timm and Tung, I-Cheng and Yang, Yafang and Li, Renkai and Yang, Jie and Weathersby, Stephen and Park, Suji and Kozina, Michael E. and Sie, Edbert J. and Wen, Haidan and Jarillo-Herrero, Pablo and Fisher, Ian R. and Wang, Xijie and Gedik, Nuh},
  journal={Nat. Phys.},
  volume={16},
  number={2},
  pages={159--163},
  year={2020},
  publisher={Nature Publishing Group UK London}
}

@article{rettig2016persistent,
  title={{Persistent order due to transiently enhanced nesting in an electronically excited charge density wave}},
  author={Rettig, L and Cort{\'e}s, R and Chu, J-H and Fisher, IR and Schmitt, F and Moore, RG and Shen, Z-X and Kirchmann, PS and Wolf, Martin and Bovensiepen, Uwe},
  journal={Nat. Commun.},
  volume={7},
  number={1},
  pages={10459},
  year={2016},
  publisher={Nature Publishing Group UK London}
}

@article{schmitt2008transient,
title = {{Transient Electronic Structure and Melting of a Charge Density Wave in TbTe$_3$}},
  author = {Schmitt, F. and Kirchmann, P. S. and Bovensiepen, U. and Moore, R. G. and Rettig, L. and Krenz, M. and Chu, J.-H. and Ru, N. and Perfetti, L. and Lu, D. H. and Wolf, M. and Fisher, I. R. and Shen, Z.-X.},
  journal={Science},
  volume={321},
  number={5896},
  pages={1649--1652},
  year={2008},
  publisher={American Association for the Advancement of Science}
}

@article{chikina2023charge,
  title={{Charge density wave generated Fermi surfaces in NdTe$_3$}},
  author={Chikina, Alla and Lund, Henriette and Bianchi, Marco and Curcio, Davide and Dalgaard, Kirstine J and Bremholm, Martin and Lei, Shiming and Singha, Ratnadwip and Schoop, Leslie M and Hofmann, Philip},
  journal={Phys. Rev. B},
  volume={107},
  number={16},
  pages={L161103},
  year={2023},
  publisher={APS}
}

@article{yumigeta2021advances,
  title={{Advances in rare-earth tritelluride quantum materials: Structure, properties, and synthesis}},
  author={Yumigeta, Kentaro and Qin, Ying and Li, Han and Blei, Mark and Attarde, Yashika and Kopas, Cameron and Tongay, Sefaattin},
  journal={Adv. Sci.},
  volume={8},
  number={12},
  pages={2004762},
  year={2021},
  publisher={Wiley Online Library}
}

@article{walmsley2020magnetic,
  title={{Magnetic breakdown and charge density wave formation: A quantum oscillation study of the rare-earth tritellurides}},
  author={Walmsley, P and Aeschlimann, S and Straquadine, JAW and Giraldo-Gallo, P and Riggs, SC and Chan, MK and McDonald, RD and Fisher, IR},
  journal={Phys. Rev. B},
  volume={102},
  number={4},
  pages={045150},
  year={2020},
  publisher={APS}
}

@article{seong2021angle,
  title={{Angle-resolved photoemission spectroscopy study of a system with a double charge density wave transition: ErTe$_3$}},
  author={Seong, Seungho and Kim, Heejung and Kim, Kyoo and Min, BI and Kwon, Yong Seung and Han, Sang Wook and Park, Byeong-Gyu and Stania, R and Seo, Yeonji and Kang, J-S},
  journal={Phys. Rev. B},
  volume={104},
  number={19},
  pages={195153},
  year={2021},
  publisher={APS}
}

@article{kang2024distinct,
  title={{Distinct charge density wave instabilities in PrTe$_n$ ($n= 2, 3$) and ErTe$_3$ investigated via ARPES and XAS}},
  author={Kang, J-S and Seong, Seungho and Lee, Eunsook and Kwon, Yong Seung and Kim, Kyoo and Kim, Junwon and Kim, Heejung and Min, BI},
  journal={Phys. Rev. Mater.},
  volume={8},
  number={8},
  pages={080301},
  year={2024},
  publisher={APS}
}

@article{raghavan2024atomic,
  title={{Atomic-scale visualization of a cascade of magnetic orders in the layered antiferromagnet GdTe$_3$}},
  author = {Raghavan, Arjun and Romanelli, Marisa and May-Mann, Julian and Aishwarya, Anuva and Aggarwal, Leena and Singh, Anisha G. and Bachmann, Maja D. and Schoop, Leslie M. and Fradkin, Eduardo and Fisher, Ian R. and Madhavan, Vidya},
  journal={npj Quantum Mater.},
  volume={9},
  number={1},
  pages={47},
  year={2024},
  publisher={Nature Publishing Group UK London}
}

@article{lei2020high,
  title={{High mobility in a van der Waals layered antiferromagnetic metal}},
  author = {Lei, Shiming and Lin, Jingjing and Jia, Yanyu and Gray, Mason and Topp, Andreas and Farahi, Gelareh and Klemenz, Sebastian and Gao, Tong and Rodolakis, Fanny and McChesney, Jessica L. and Ast, Christian R. and Yazdani, Ali and Burch, Kenneth S. and Wu, Sanfeng and Ong, Nai Phuan and Schoop, Leslie M.},
  journal={Sci. Adv.},
  volume={6},
  number={6},
  pages={eaay6407},
  year={2020},
  publisher={American Association for the Advancement of Science}
}

@article{ru2008magnetic,
  title={{Magnetic properties of the charge density wave compounds $R$Te$_3$ ($R=\,$Y, La, Ce, Pr, Nd, Sm, Gd, Tb, Dy, Ho, Er, and Tm)}},
  author={Ru, N and Chu, J-H and Fisher, IR},
  journal={Phys. Rev. B},
  volume={78},
  number={1},
  pages={012410},
  year={2008},
  publisher={APS}
}

@article{Ru2008cdw,
  title = {{Effect of chemical pressure on the charge density wave transition in rare-earth tritellurides $R{\mathrm{Te}}_{3}$}},
  author = {Ru, N. and Condron, C. L. and Margulis, G. Y. and Shin, K. Y. and Laverock, J. and Dugdale, S. B. and Toney, M. F. and Fisher, I. R.},
  journal = {Phys. Rev. B},
  volume = {77},
  issue = {3},
  pages = {035114},
  numpages = {9},
  year = {2008},
  month = {Jan},
  publisher = {American Physical Society},
  doi = {10.1103/PhysRevB.77.035114},
  url = {https://link.aps.org/doi/10.1103/PhysRevB.77.035114}
}

@article{Nakajima2024PONTA,
  author  = {Nakajima, Taro and Saito, Hiraku and Kobayashi, Naoki and Kawasaki, Takuro and Nakamura, Tatsuya and Kawano-Furukawa, Hazuki and Asai, Shinichiro and Masuda, Takatsugu},
  title   = {{Polarized and Unpolarized Neutron Scattering for Magnetic Materials at the Triple-axis Spectrometer {PONTA} in {JRR-3}}},
  journal = {J. Phys. Soc. Jpn.},
  year    = {2024},
  volume  = {93},
  number  = {9},
  pages   = {091002},
  doi     = {10.7566/JPSJ.93.091002},
  url     = {https://doi.org/10.7566/JPSJ.93.091002}
}

@article{PhysRevB.90.085105,
  title = {{Coexistence and competition of multiple charge-density-wave orders in rare-earth tritellurides}},
  author = {Hu, B. F. and Cheng, B. and Yuan, R. H. and Dong, T. and Wang, N. L.},
  journal = {Phys. Rev. B},
  volume = {90},
  issue = {8},
  pages = {085105},
  numpages = {7},
  year = {2014},
  month = {Aug},
  publisher = {American Physical Society},
  doi = {10.1103/PhysRevB.90.085105},
  url = {https://link.aps.org/doi/10.1103/PhysRevB.90.085105}
}

@article{iyeiri2003magnetic,
  title={{Magnetic properties of rare-earth metal tritellurides $R$Te$_3$ ($R=\,$Ce, Pr, Nd, Gd, Dy)}},
  author={Iyeiri, Yuji and Okumura, Teppei and Michioka, Chishiro and Suzuki, Kazuya},
  journal={Phys. Rev. B},
  volume={67},
  number={14},
  pages={144417},
  year={2003},
  publisher={APS}
}

@article{shamova2025comparative,
  title={{Comparative study of terbium tellurides Tb$_2$Te$_5$ and TbTe$_3$}},
  author = {Shamova, I. and Sun, J.-Y. and Chang, X.-Y. and Popova, V. and Chareev, D. and Shvanskaya, L. and Ksenofontov, D. and Vorobyova, A. and Lyssenko, K. and Demidov, A. and Yeh, H.-Y. and Tzeng, W.-Y. and Lin, J.-Y. and Luo, C.-W. and Monceau, P. and Pachaud, E. and Lorenzo, E. and Sinchenko, A. and Vasiliev, A. and Volkova, O.},
  journal={Phys. Rev. B},
  volume={112},
  number={9},
  pages={094434},
  year={2025},
  publisher={APS}
}

@inproceedings{deguchi2009magnetic,
  title={{Magnetic order of rare-earth tritelluride CeTe$_3$ at low temperature}},
  author={Deguchi, K and Okada, T and Chen, GF and Ban, S and Aso, N and Sato, NK},
  booktitle={Journal of Physics: Conference Series},
  volume={150},
  issue={4},
  pages={042023},
  year={2009},
  organization={IOP Publishing}
}

@article{Pfuner2010,
  author    = {Pfuner, F. and Lerch, P. and Chu, J.-H. and Kuo, H.-H. and Fisher, I. R. and Degiorgi, L.},
  title     = {{Temperature dependence of the excitation spectrum in the charge-density-wave {ErTe}$_3$ and {HoTe}$_3$ systems}},
  journal   = {Phys. Rev. B},
  volume    = {81},
  number    = {19},
  pages     = {195110},
  numpages  = {7},
  year      = {2010},
  month     = {May},
  publisher = {American Physical Society},
  doi       = {10.1103/PhysRevB.81.195110},
  url       = {https://doi.org/10.1103/PhysRevB.81.195110}
}

@article{PhysRevB.77.235104,
  title = {{Angle-resolved photoemission study of the evolution of band structure and charge density wave properties in $R{\text{Te}}_{3}$ ($R=\text{Y}$, La, Ce, Sm, Gd, Tb, and Dy)}},
  author = {Brouet, V. and Yang, W. L. and Zhou, X. J. and Hussain, Z. and Moore, R. G. and He, R. and Lu, D. H. and Shen, Z. X. and Laverock, J. and Dugdale, S. B. and Ru, N. and Fisher, I. R.},
  journal = {Phys. Rev. B},
  volume = {77},
  issue = {23},
  pages = {235104},
  numpages = {16},
  year = {2008},
  month = {Jun},
  publisher = {American Physical Society},
  doi = {10.1103/PhysRevB.77.235104},
  url = {https://link.aps.org/doi/10.1103/PhysRevB.77.235104}
}

@article{PhysRevB.77.035114,
  title = {{Effect of chemical pressure on the charge density wave transition in rare-earth tritellurides $R{\mathrm{Te}}_{3}$}},
  author = {Ru, N. and Condron, C. L. and Margulis, G. Y. and Shin, K. Y. and Laverock, J. and Dugdale, S. B. and Toney, M. F. and Fisher, I. R.},
  journal = {Phys. Rev. B},
  volume = {77},
  issue = {3},
  pages = {035114},
  numpages = {9},
  year = {2008},
  month = {Jan},
  publisher = {American Physical Society},
  doi = {10.1103/PhysRevB.77.035114},
  url = {https://link.aps.org/doi/10.1103/PhysRevB.77.035114}
}

@article{akatsuka2024non,
  title={{Non-coplanar helimagnetism in the layered van-der-Waals metal DyTe$_3$}},
  author = {Akatsuka, Shun and Esser, Sebastian and Okumura, Shun and Yambe, Ryota and Yamada, Rinsuke and Hirschmann, Moritz M. and Aji, Seno and White, Jonathan S. and Gao, Shang and Onuki, Yoshichika and Arima, Taka-hisa and Nakajima, Taro and Hirschberger, Max},
  journal={Nat. Commun.},
  volume={15},
  number={1},
  pages={4291},
  year={2024},
  publisher={Nature Publishing Group UK London}
}

@article{chillal2020strongly,
  title={{Strongly coupled charge, orbital, and spin order in TbTe$_3$}},
  author = {Chillal, S. and Schierle, E. and Weschke, E. and Yokaichiya, F. and Hoffmann, J.-U. and Volkova, O. S. and Vasiliev, A. N. and Sinchenko, A. A. and Lejay, P. and Hadj-Azzem, A. and Monceau, P. and Lake, B.},
  journal={Phys. Rev. B},
  volume={102},
  number={24},
  pages={241110},
  year={2020},
  publisher={APS}
}

@inproceedings{hanai2024arim,
    title = {{ARIM-mdx Data System: Towards a Nationwide Data Platform for Materials Science}},
    author = {Hanai, Masatoshi and Ishikawa, Ryo and Kawamura, Mitsuaki and Ohnishi, Masato and Takenaka, Norio and Nakamura, Kou and Matsumura, Daiju and 
        Fujikawa, Seiji and Sakamoto, Hiroki and Ochiai, Yukinori and Okane, Tetsuo and Kuroki, Shin-Ichiro and Yamada, Atsuo and Suzumura, Toyotaro and
        Taura, Kenjiro and Mita, Yoshio and Shibata, Naoya and Ikuhara, Yuichi},
    booktitle = {Proceedings of 2024 IEEE International Conference on Big Data (BigData)},
    pages={2326-2333},
    year = {2024}
}

@article{PhysRevB.91.205114,
  title = {{Pressure dependence of the charge-density-wave and superconducting states in ${\text{GdTe}}_{3}, {\text{TbTe}}_{3}$, and ${\text{DyTe}}_{3}$}},
  author = {Zocco, D. A. and Hamlin, J. J. and Grube, K. and Chu, J.-H. and Kuo, H.-H. and Fisher, I. R. and Maple, M. B.},
  journal = {Phys. Rev. B},
  volume = {91},
  issue = {20},
  pages = {205114},
  numpages = {7},
  year = {2015},
  month = {May},
  publisher = {American Physical Society},
  doi = {10.1103/PhysRevB.91.205114},
  url = {https://link.aps.org/doi/10.1103/PhysRevB.91.205114}
}

@misc{SI,
  note = {See Supplementary Information at
    (URL to be inserted by publisher).},
}

\end{document}